# OPTICAL AND NANOSTRUCTURAL INSIGHTS OF OBLIQUE ANGLE DEPOSITED LAYERS APPLIED FOR PHOTONIC COATINGS.

*Florian Maudet[1,4*], Bertrand Lacroix[2,3], Antonio J. Santos[2,3], Fabien Paumier[4*], Maxime Paraillous[5], Simon Hurand[4], Alan Corvisier[4], Cecile Marsal[4], Baptiste Giroire[5], Cyril Dupeyrat[5], Rafael García[2,3], Francisco M. Morales[2,3] and Thierry Girardeau[4]*

*[1] Institute Functional thin film oxides for energy-efficient future information technology, Helmholtz-Zentrum Berlin für Materialien und Energie Hahn-Meitner-Platz 1, 14109, Berlin, Germany*
*[2] Department of Materials Science and Metallurgic Engineering, and Inorganic Chemistry, Faculty of Sciences, University of Cádiz, Spain.*
*[3] IMEYMAT: Institute of Research on Electron Microscopy and Materials of the University of Cádiz, Spain.*
*[4] Institut Pprime, UPR 3346 CNRS-Université de Poitiers-ENSMA, SP2MI, 86962 Futuroscope-Chasseneuil cedex, France*
*[5] Safran Electronics and Defense, 26 avenue des Hauts de la Chaume, 86280 Saint-Benoît, France*

Oblique angle deposition (OAD) is a nanostructuration method widely used to tune the optical properties of thin films. The introduction of porosity controlled by the deposition angle is used to develop the architecture of each layer and stack that enable modifying and optimizing the optical properties of the constituent layers for optimal design. However, optical properties of these nanostructured layers may differ greatly from those of dense layers due to the presence of anisotropy, refractive index gradient and scattering. This work focuses on OAD layers based on a reference photonic material such as $SiO_2$ and it aims at taking into account all these effects in the description of the optical response. For that, the nanostructure has been analyzed with a complete SEM study and key parameters like the porosity gradient profile and aspect ratio of the nanocolumns were extracted. The samples were then characterized by generalized ellipsometry to evaluate the influence of morphological anisotropy and porosity gradient on the optical response of the films. Based on this microstructural study, an original optical model is presented to fit the features of new optical properties. A reliable correspondence is observed between the optical model parameters and the microstructure characteristics like the column angle and the porosity gradient. This demonstrates that such complex microstructural parameters can be easily accessed solely from optical measurements. All the work has enabled us to develop a two-layer anti-reflective coating that already demonstrate high level of transmission.

*florian.maudet@helmholtz-berlin.de

*fabien.paumier@univ-poitiers.fr

# I. INTRODUCTION

Oblique angle deposition (OAD) is a versatile method to obtain complex morphological nanostructures for a great variety of materials [1]. By changing the deposition angle, it is possible to take advantage of the self-shadowing effect to control the porosity rate at the nanoscale [2]. These OAD thin films, are generally made of slanted nanocolumns, that will exhibit tunable optical properties that differs from their bulk counterpart due to the presence of porosity and morphological anisotropy, making OAD very interesting for various applications like antireflective coatings, Bragg reflectors, solar cells or polarized optics [3]–[9]. However, both the anisotropy and porosity of the OAD films strongly depend on the growth mechanisms that may vary with the deposited material as well as the deposition parameters [10], [11]. To engineer the optical properties of OAD thin films, it is thus necessary not only to understand and control the growth of a specific material in relation with the deposition parameters, but also to accurately model the optical response of these complex structures. In most cases, the conventional Bruggeman effective medium approximation (BEMA) is used to describe the optical properties of mesoporous layers that are generally considered as a mixture of dense material and air [3], [12], [13]. However, this simple interpretation does not take into account the morphological anisotropy specific of OAD layers, and more advanced models like the Anisotropic BEMA (ABEMA) should be used instead [14]. Moreover, as demonstrated from microstructural and simulation studies, the porosity and morphological anisotropy of OAD layers evolves during the growth [4], [15]. Integrating anisotropy, porosity and their evolution in the optical description of thin films thus involves using a multilayered model with graded optical properties instead of a single homogenous effective layer as it is usually done. This approach can lead to a more accurate description of the optical response that provides information on the morphological parameters of the layers.

In this article, we aim at knowing to which extent such an optical model can describe the observed morphological properties of OAD thin films. We studied the case of $SiO_2$ because it is a reference photonic material and it can be nanostructured by OAD to obtain very low refractive index that is needed for a high-performance antireflective coating [5]. We demonstrate the interest of this knowledge by developing a $SiO_2$ antireflective bilayer presenting not only a low reflectance but also a high transmittance over a large band of wavelength.

We present herein first a SEM based microstructural study of a series of $SiO_2$ deposited by OAD made to extract key morphological features to help building an optical model. These microstructural observations and their possible physical origin are then discussed in regard of the existing literature. From those observations, an optical model is proposed and is then applied to model generalized ellipsometry and spectrophotometry measurements of the prepared films. The parameters extracted from the optical model are then compared to the morphological parameters observed for the $SiO_2$ series. To illustrate the optical model and its sensitivity to optical measurements a python code that calculates the ellipsometric and spectrophotometric response of the optical model is presented in Supplementary Material 1 of this article [16]. To conclude, this model is used to optimize and describe the optical response of a $SiO_2$ antireflective bilayer, the morphological parameters extracted from the optical model are compared to electron tomography observation.

# II. EXPERIMENTAL DETAILS

The experimental setup used for the oblique angle deposition consists in a vacuum chamber (base pressure $2.10^{-6}$ mbar) equipped with an electron-beam evaporator. $SiO_2$ OAD layers were deposited at various angles of incidence from 0° to 85° by e-beam evaporation method using a $SiO_2$ crucible (Umicore© purity>99.99%) on 1 mm thick, 1-inch diameter BK7 glass substrates. For each deposition, a (001) silicon substrate was also placed, with one <110> crystallographic axis perpendicular to the incoming deposition flux for further analyses by transmission electron microscopy (TEM) and spectrometric ellipsometry studies. A quartz crystal monitor was used to control the thickness and ensure a nominal deposition rate of 10 Å/s at normal incidence. The depositions were all performed at room temperature. After calibration, a $SiO_2$ bilayer was deposited by OAD. Since our deposition set-up is not equipped with an in-situ adjustable orientation sample-holder, the bilayer was made in two steps with an azimuthal rotation of Φ=180° in between on silicon glass (BK7) substrate. For the first step, a $SiO_2$ layer was deposited at an angle of 65° with the condition of elaboration previously described. For the second step, the chamber was opened to modify the angle of incidence to a value of 85°. The second layer of $SiO_2$ was then evaporated and deposited onto the first OAD layer. This whole process was repeated a second time to coat the BK7 substrate on both sides.

The microstructural study was performed by scanning electron microscopy (SEM) using a FEG 7001F-TTLS JEOL microscope at an accelerating voltage of 30kV. To extract statistical data from the micrographs (Figure 1), different types of filters were used to i) correct and improve the quality of contrasts, ii) highlight the contours of the nano-columns, iii) reduce image noise. Advanced image processing was then implemented. The $SiO_2$ columns were isolated by locally segmenting the image. After this segmentation, the void appears in dark contrast and the condensed matter in light contrast, which makes possible the easy identification of the columns or column agglomerates (Figure 1.b and c). Finally, we carried out a statistical analysis by adjusting an ellipse as closely as possible to the contour of the column or agglomerate; each ellipse being characterized by its area, ellipticity and orientation (Figure 1.d).

To access the optical properties of monolayers and bilayers, spectrophotometry measurements were carried out for each sample. Transmittance and reflectance spectra were recorded with a Carry 5000 Varian spectrophotometer. Transmittance and reflectance acquisitions were repeated 4 times for each sample giving a repeatability of 0.1%. To avoid problems coming from the anisotropic behavior of the samples, measurements were made with a light polarized in the direction perpendicular to the direction of deposition, *i.e.* along the elongation axis of the nanocolumns. Indeed, as we will see later, ellipsometry study of OAD samples show anisotropic character such as non-zero off diagonal coefficients per block of the Mueller matrix. Consequently, a generalized ellipsometry characterization was used to improve the optical model. This generalized ellipsometry characterization was performed with a Woolam M-2000, and the measurements were carried out with an incidence angle going from 55° to 80° with a 5° step for each azimuthal rotation angle going from 0° to 180° with a 5° step.

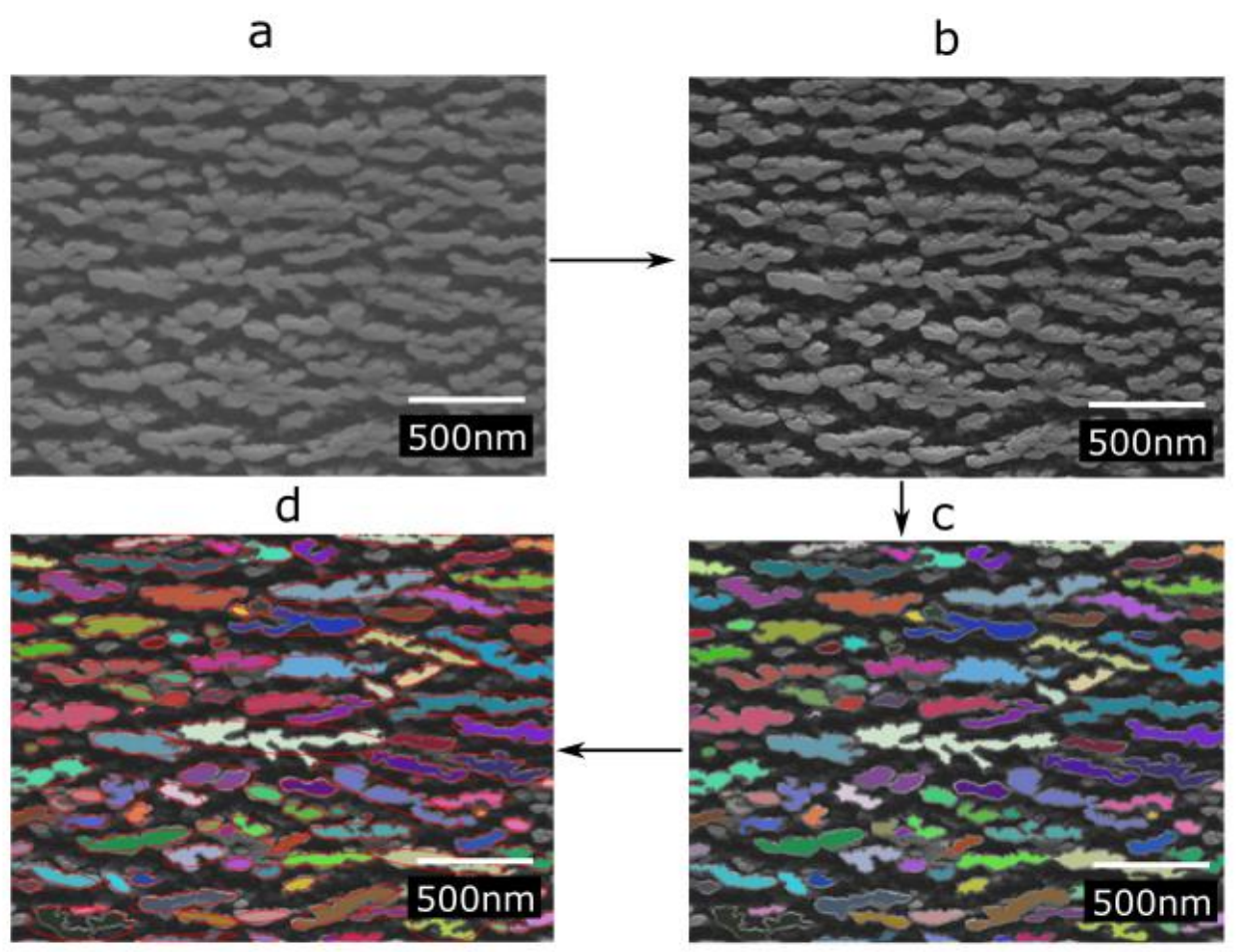

Figure 1 : SEM micrographs for different processing steps, (a) raw picture, (b) after cleaning with different filters, (c) after segmentation and column detection, (d) micrograph after column adjustment with ellipses

The nanostructure of the OAD samples was analyzed by TEM experiments at an acceleration voltage of 200 kV. Bright-field TEM images were recorded in a JEOL 2010F microscope. The 3D morphology of the OAD $SiO_2$ bilayer deposited on the substrate was extracted from electron tomography in a FEI Titan Cubed Themis 60-300 microscope, using the high-angle annular dark-field scanning TEM (HAADF-STEM) imaging mode. For that purpose, a dedicated tomography holder operated by the FEI Xplore3D software was used to acquire tilt series every 2° from -60° to +60°. The FEI Inspect 3D software was then employed to align the projections using cross-correlation methods, and to perform reconstruction into a 3D volume using the conventional simultaneous iterative reconstruction technique (SIRT) with 30 iterations. The FEI Avizo program was used for 3D visualization and segmentation of the reconstructed volume. Specimen for TEM were prepared by soft and flat mechanical polishing of the two faces using a tripod apparatus (Model 590 Tripod Polisher) in order to control finely the thinning down to a few microns. This step was followed by a short $Ar^+$-ion milling step in a Gatan PIPS system (using an energy of 3.5 keV and an incidence angle of +/-7º). This procedure provides large and homogeneous electron transparent areas, which is suitable to limit the geometrical shadowing of the specimen at large tilt angles of the holder. Prior to insertion into the microscope, the specimen was cleaned in an $Ar/O_2$ plasma in order to remove the hydrocarbon contamination.

## III. RESULTS AND DISCUSSION

### 1.Microstructural characterization

In order to build an accurate optical model, we investigated at first the microstructural properties of the samples. As it is expected, the column angle $\beta_{col}$ of our $SiO_2$ OAD layers increases with the deposition angle α, from $\beta_{col}$=24º for α=65° up to $\beta_{col}$=38º for α=90° (Figure 2). These results matched the ones observed and predicted by Alvarez et al. for $SiO_2$ [10]. On the other hand, we observe that the growth of some columns has been stopped within the films due a competitive growth process wherein the largest columns are shading the smallest ones. This phenomenon contributes to the evolution of porosity as function of the depth inside the film [17].

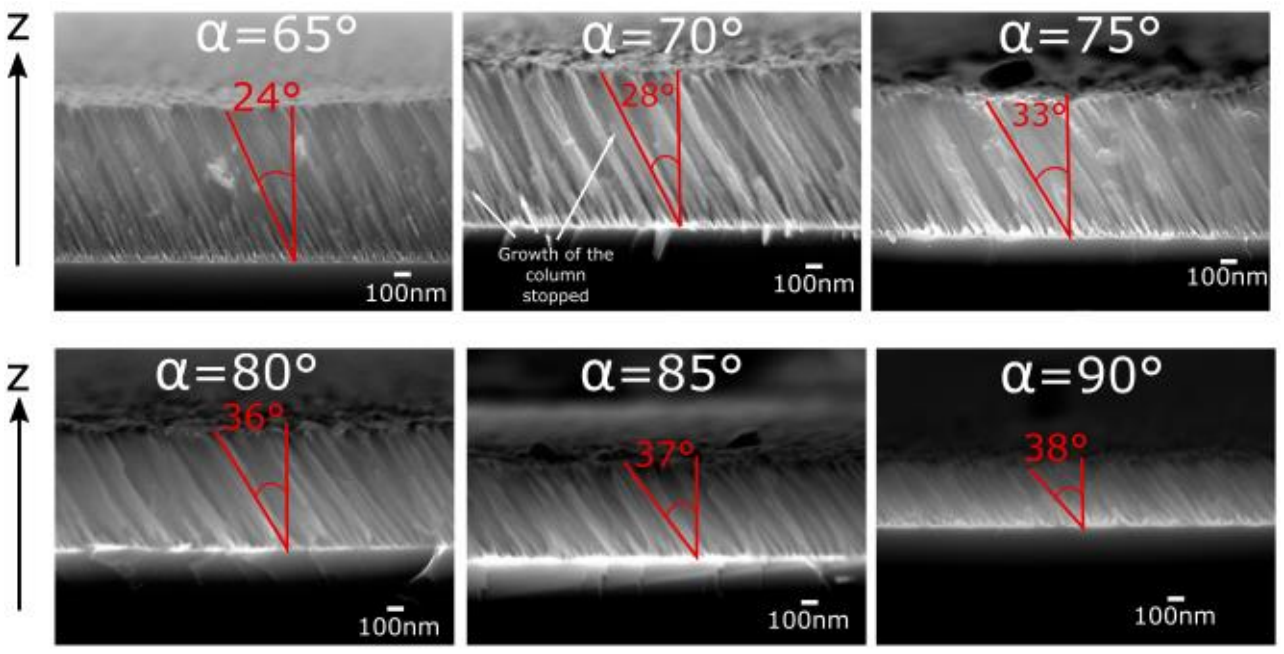

Figure 2 Evolution of the column angle $\beta_{col}$ with the deposition angle α for a $SiO_2$ series

As it is hard to get information on anisotropy or porosity evolution of the thin film solely from cross-sectional observations, SEM images were also captured in the direction of the columns. Under these conditions, four SEM micrographs of $SiO_2$ deposited at 85° with different thicknesses have been performed: 76 nm, 153 nm, 317 nm, 630 nm (Figure 3). We qualitatively observe, with the increase in thickness, a coalescence of the columns leading to an anisotropic enlargement of the structures. These observations indicate increasingly anisotropic structures with the film thickness, in accordance with those made by Vick et al. and C. Lopez-Santos et al. [4], [18].

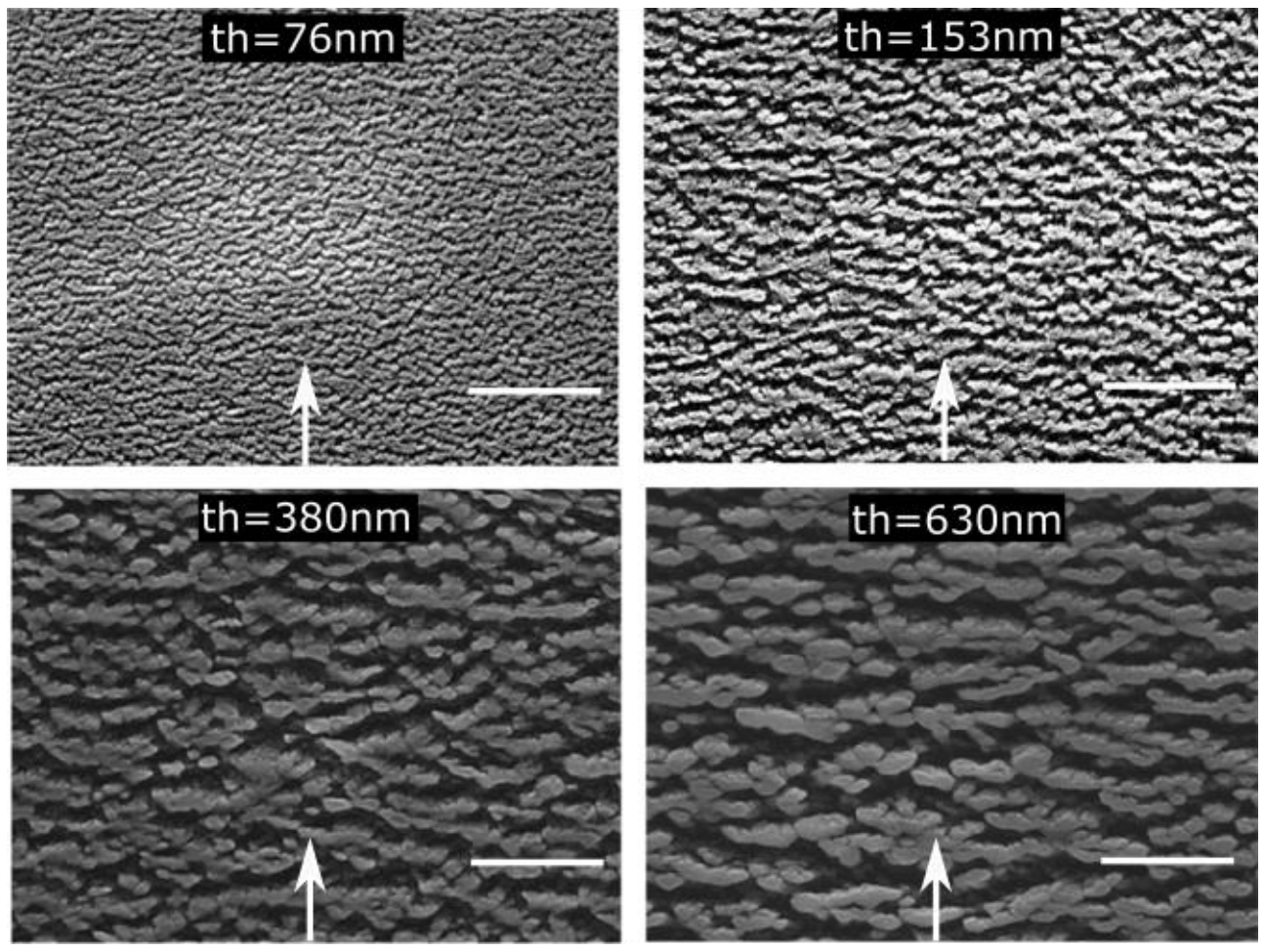

Figure 3 SEM micrographs recorded along the column direction of the $SiO_2$ sample deposited by OAD at α=85°, for different film thicknesses (indicated on top of each image). The arrows indicate the direction of the incoming species during the deposition process, the scale bar is 500nm.

To be more quantitative, we present on Figure 4.a the average area evolution of the columns agglomerates for different thicknesses using the method described in Part II (for more information see Supplementary Material 2 (colored image)). The linear growth of the average area of the columns section as a function of deposition thickness indicates a columnar broadening of structures deposited by OAD, which has been reported in other works [4], [19], [20]. This effect should lead to a decrease in porosity with thickness. The competitive growth phenomenon mentioned previously is expected to have an opposite effect, leading instead to a porosity increase with thickness. According to Figure 4.b, the observed large increase in surface porosity with the thickness of the thin film actually suggests that the latter phenomenon is dominant, at least over the thickness range studied. Furthermore, it can be noted that the porosity increases faster for the first part of the growth process and slows down for the second part. This suggests that the early stage of the growth is dominated by the competitive growth process when for the second part this mechanism is balanced by the column broadening process.

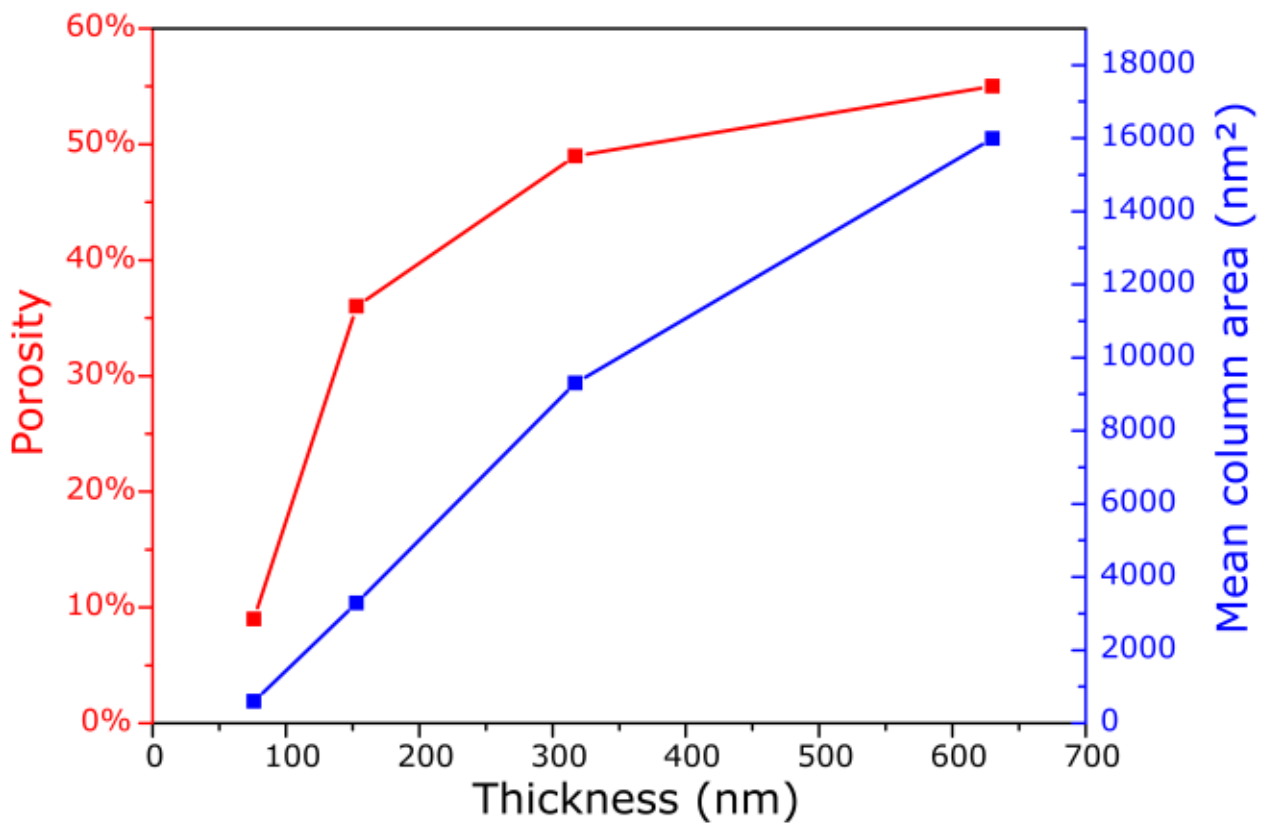

Figure 4 Evolution of the surface porosity (red) and of the mean column area (blue) for a $SiO_2$ sample deposited at α= 85° for different thicknesses

Using the information extracted from the ellipses approximating the nanostructures observed on SEM images, the evolution of the anisotropic character as a function of the film thickness can also be highlighted in Figure 5. To describe this anisotropy, a coordinate system (x', y', z') linked to the columns is introduced, tilted from β angle along y axis of the laboratory system (x, y, z) (Figure 6). These ellipses are characterized by their major axis $l_b$ along y'=y axis of Figure 6, their minor axis $l_a$ along x' axis and the angle γ indicating the orientation of the ellipses with respect to the horizontal axis of Figure 3. The evolution of the average length of these axes (Figure 5.a) clearly points out the development of morphological anisotropy as a function of thickness. Indeed, there is a slight increase in the average length of the short axis with thickness, while the average length of the long axis increases more rapidly. This reflects an anisotropic elongation of the structures during the film growth, as evidenced by the increase of the ratio $l_b/l_a$.

On the other hand, the increase in anisotropy is accompanied by a progressive alignment of the ellipses. This behavior is materialized by the probability distribution of the orientation of the ellipses increasingly centered on the horizontal axis as the thickness increases Figure 5.b).

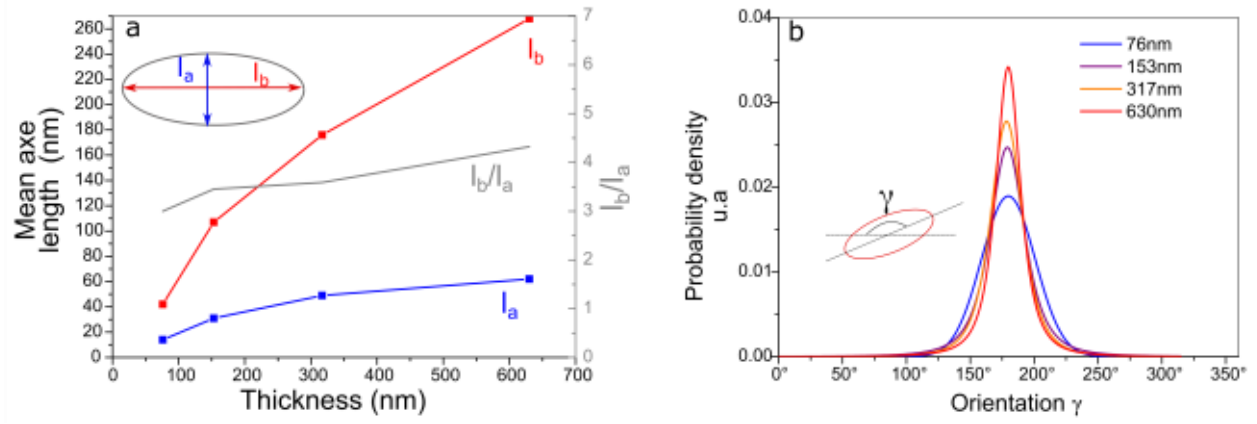

Figure 5 (a) Evolution of the morphological parameters of $SiO_2$ nanocolumns (assimilated to ellipses with their minor and major axes Ia and Ib, respectively) as a function of the thickness of the (b) Evolution of the probability density of the orientation γ with respect to the horizontal axis in Figure 3.

## 2. Microstructural synthesis and discussion

To sum up, the microstructural characterizations have highlighted four morphological characteristics important for deepening the comprehension of the optical properties of OAD-formed thin films: (i) an increase of the columns angle $\beta_{col}$ in respect to the deposition angle α, (ii) an increase in porosity rate during the growth, (iii) a broadening of the column section for an increasing thickness, (iv) an anisotropic growth of the nanostructures.

Thus, OAD deposition involves different competing growth processes. The shadowing effect favors the growth of the largest columns and the extinction of the growth of the smallest columns [17]. As highlighted in the work of F. Nita *et al.*[15], this process is responsible for the increase in porosity rate during thin film growth. These observations characterize Phase I of the diagram presented in Figure 6. However, the increase in porosity rate during growth is compensated by the columnar broadening mechanism, Phase II in Figure 6. For low incident species energies, as it is the case for evaporation at room temperature, Alvarez *et al.* highlighted the importance of dispersion (Δα) of species incidence to explain the increase in column cross section [10]. Indeed, species deposited on the columns in different directions will not only contribute to increasing their length but also to their expansion. Thus the increase in angular aperture is accompanied by an acceleration of the columnar broadening [21]. However, this purely geometric explanation alone cannot explain the differences observed between materials.

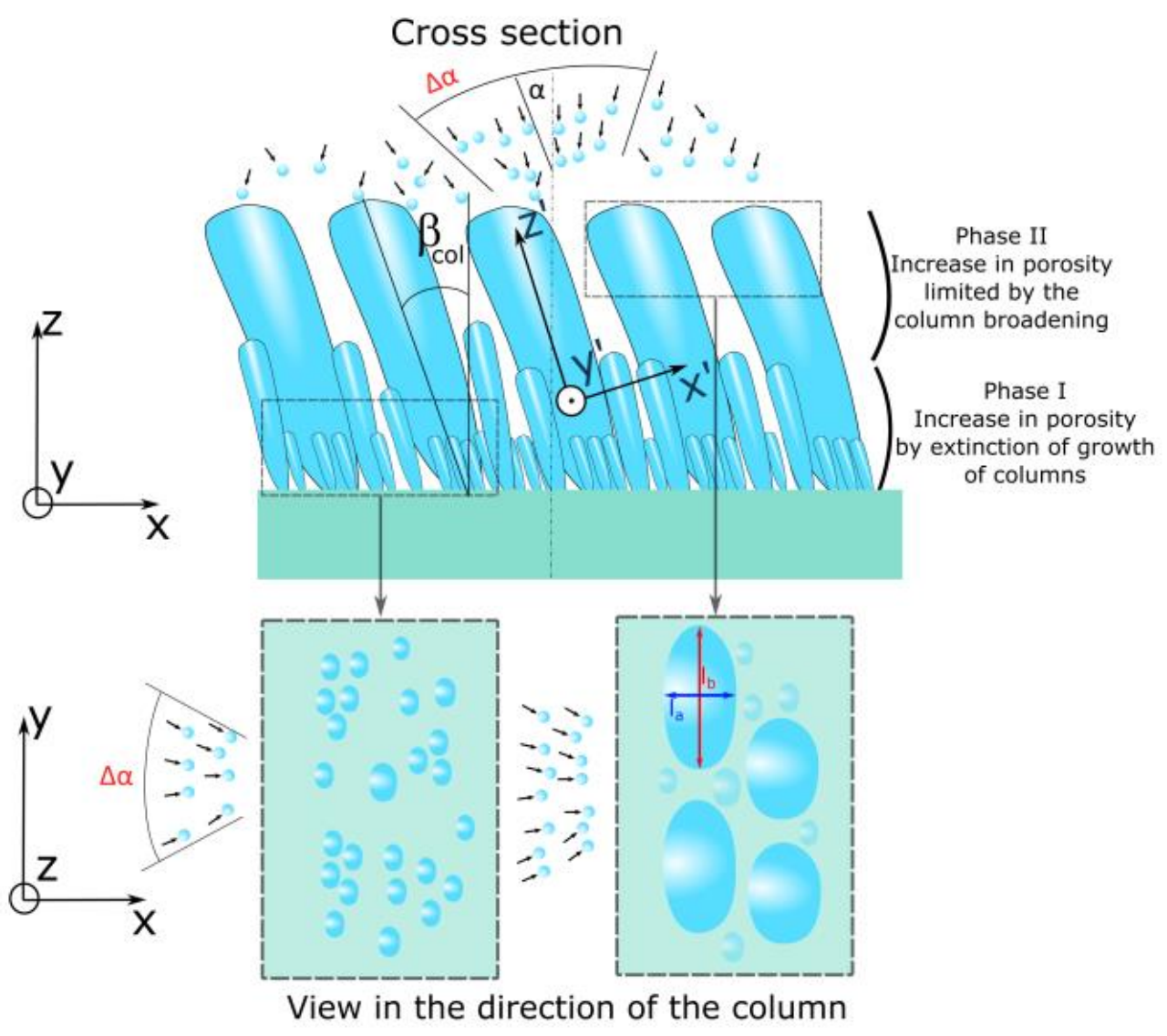

Figure 6 Schematic view of the columns broadening occurring during OAD growth (a) cross section highlighting the different regime of porosity rate (b) plane view. The coordinate system (x', y', z') linked to columns is introduced for a clearer description of the anisotropy.

C. Lopez-Santos *et al.* have highlighted the phenomenon of species captured passing near the columns [4]. By considering different capture radii (named the sticking coefficient in the reference) the authors were able to explain the differences in growth behavior specific to different materials[10]. They showed that this capture effect favors columnar enlargement and therefore the coalescence of the columns. This effect is also at the origin of anisotropic growth, for species with a large catch radius the expansion of the section is favored. This broadening favors the coalescence of the columns between them in the direction orthogonal to the flow because there is no shading effect. The resulting structure is an elongated ellipse orthogonally to the flow [4]. Our observations are thus in line with the OAD growth model of $SiO_2$.

## 3. Optical model

Usually OAD deposited layers are simulated as one thin heterogenous layer composed of porosity and dense material. This optical description is however very basic and it will fail to consider the porosity evolution as well as the anisotropic morphology that was pointed out in the previous section. In order to sense the influence of these crucial geometrical factors from the optical point of view, generalized ellipsometry characterization is used to measure the Mueller matrix of the samples. The simulation of the experimental data, which requires the use of a model taking into account the anisotropy, gives access to a lot of information. The model we used is the Anisotropic Bruggeman Effective Medium Approximation (ABEMA) model. This model determines the porosity ratio, the angle of inclination $\beta_{col}$ of the optical axis (equivalent in our case to the angle of the columns) as well as the depolarization factors La, Lb and Lc at the origin of anisotropy associated with the shape factors of nanostructures along x', y' and z' directions respectively.

To consider the porosity evolution along the thickness of the films (z axis) evidenced by the microscopic observations, a porosity gradient was applied by dividing the thin film into 20 layers of variable porosity. To best match the evolution shown in Figure 4.b, this gradient was decomposed into two linear segments. A first segment adapted to the evolution of the porosity over 10% of the initial thickness of the film and a second segment adapted to the evolution of the porosity of the remaining thickness. The ordinate at the origin of the first segment and the slopes of the two segments are therefore the adjustable parameters of the porosity gradient. Although the microstructural study pointed out a morphological increase of the anisotropy as function of z, this was not implemented in the model considered in this article. We therefore assumed fixed depolarization factors representing the morphological anisotropy for the entire thin film. A detailed discussion on this topic can be found in Supplementary Material 3.

The ABEMA model is considered to be valid for nanostructure dimensions that are much smaller than the probing wavelength, typically $r/\lambda<0.1$ where r is the radius of a spherical porosity and $\lambda$ the wavelength. For inclusion dimensions that do not fulfil this criterion, diffusion mechanisms must be taken into account [12], [22]. To account for the possible presence of diffusion close to the UV spectral range, an intensity attenuation was applied according to Urbach's rule [23]. Spectrophotometric measurements can only measure optical losses without distinguishing between scattering and absorption. The absorption of dense $SiO_2$ embedded in OAD nanostructure is assumed equal to the one evidenced on a dense sample deposited at α=0°. In this way, possible additional optical losses evidenced by OAD nanostructures are assumed coming from diffusion. The validity of this hypothesis is discussed in Supplementary Material 4.

In summary, the adjustable parameters of the optical model are therefore: the thickness of the OAD layer e; the volume fraction of air (porosity) near the substrate (z=0), at 10% of the film thickness and on the surface of the layer (z=e); the amplitude and slope of Urbach absorption (that in fact originates from diffusion); The depolarization coefficient Lc and the ratio of the depolarization coefficients La/Lb; the columns angle (optical axis) $\beta_{opt}$ in respect with the y axis and Φ the azimuth angle of these columns in respect with the z axis. To illustrate the model and its sensitivity to optical measurements a python code that allows the calculation of all optical simulation is presented in form of a Jupyter notebook on Supplementary Material 1.

## 4. Optical simulation of experimental OAD layers

As an example, we present on Figure 7 the advanced optical characterization for an $SiO_2$ OAD thin film deposited at α=70°. The generalized ellipsometry measurements (7.c.) have been acquired with an incidence angle θ=65° and with the optimal azimuthal angle for anisotropic contrast: φ=135° from x and y axis. These ellipsometry measurements are performed on $SiO_2$ layers deposited and silicon substrates when the layers analyzed by spectrophotometry are deposited on glass. The same model is applied for both $SiO_2$ OAD layers with the hypothesis that the substrate doesn't affect their growth behavior. The model shown in Figure 7.d summarizes the results.

The very good agreement between the simulations and the numerous experimental measurements attests to the quality and robustness of the model used. The information about the anisotropy of the sample carried by the off diagonal block coefficients of the Mueller matrix makes possible the determination of the optical axis angle of the OAD layers: $\beta_{opt}$=25° for this thin film deposited at α=70°. From the depolarization coefficients $L_a$, $L_b$ and $L_c$ it is possible to determine the characteristic shape factor, considering a medium composed of aligned ellipsoids with the average aspect ratio of the $SiO_2$ OAD ellipsoids nanostructures [24]. In particular, the aspect ratio of the ellipses is in this case $r = l_b/l_a = 2.31$.

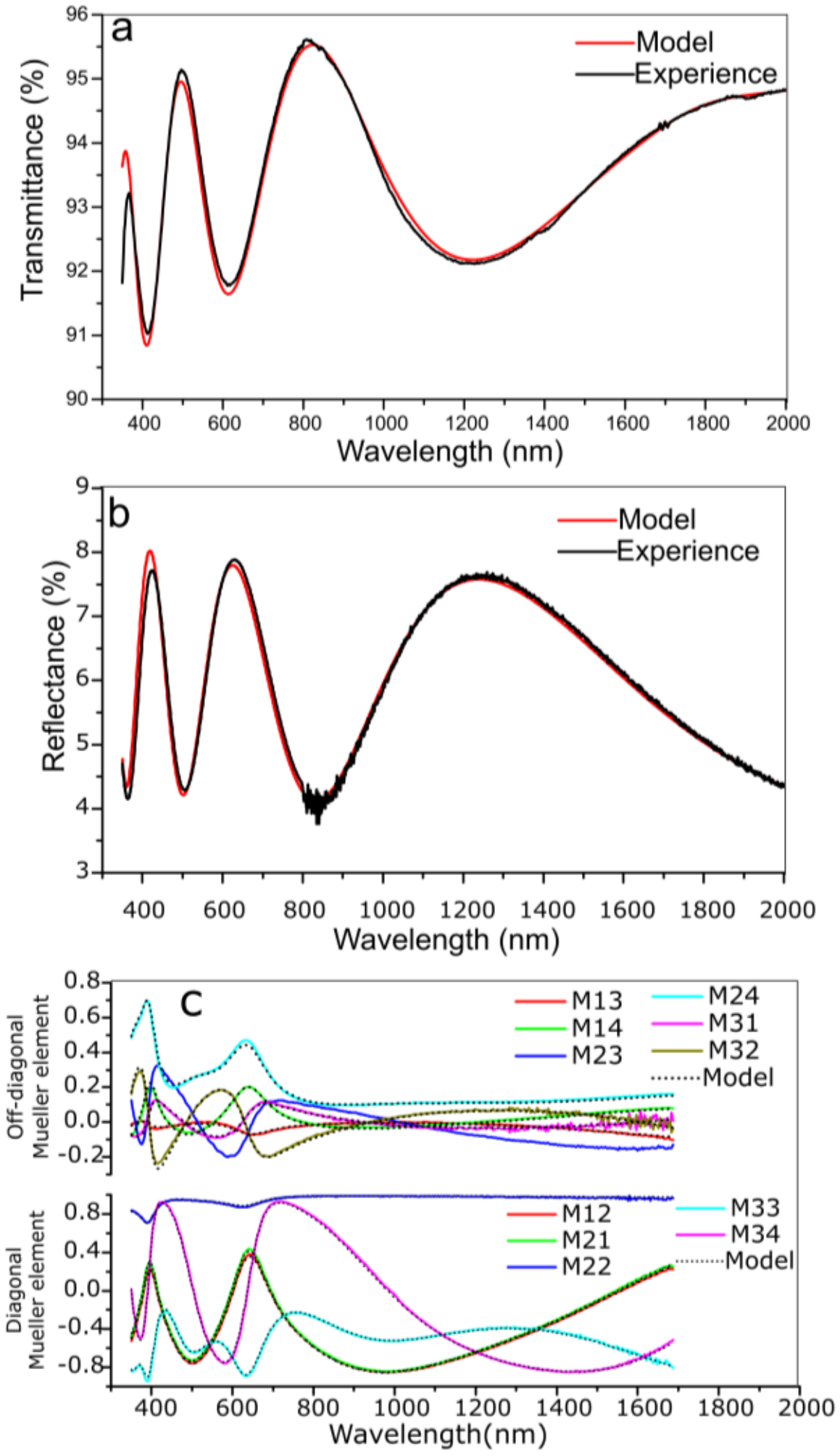

Figure 7 Experimental and simulated optical spectra. (a) Normal transmission. (b) Quasi-normal reflection. (c) Off diagonal coefficients per block of the Mueller matrix determined by generalized ellipsometry at θ=65° and φ=135° or an $SiO_2$ OAD thin film deposited at α=70°

Since the model is anisotropic, the effective refractive index differs according to the three directions of the sample reference (x', y' and z'). These indices also change between the substrate and the interface with the surrounding environment due to the porosity gradient. This gradient extracted from the optical simulation is given in Figure 8.a. The evolution of the resulting refractive indices (Figure 8.b) is illustrated by their values at the ends of the OAD layer: positions marked by the arrows (z=0 and z=e) in Figure 8.a.

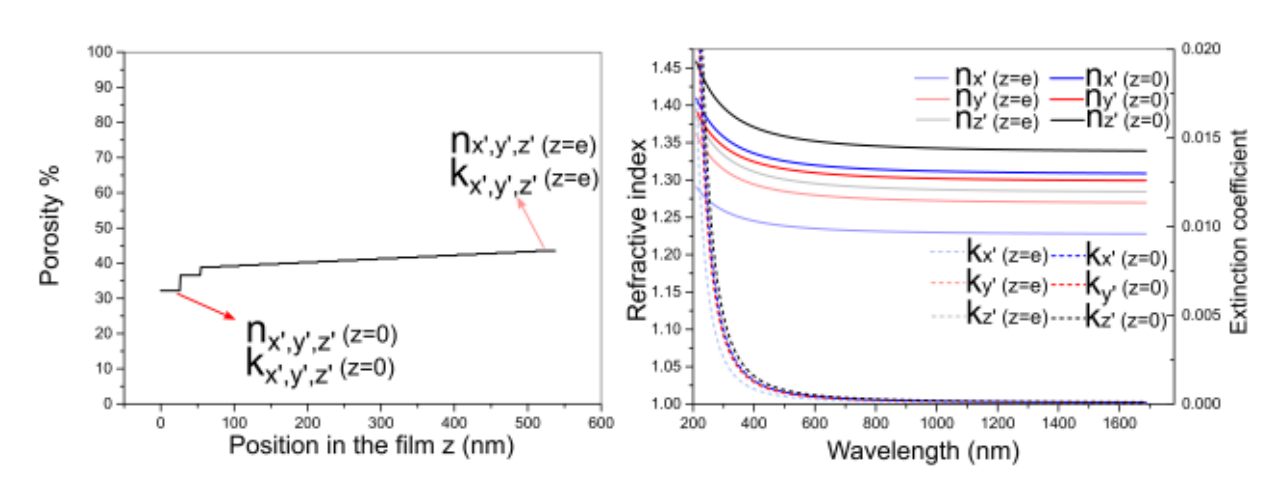

Figure 8 Parameters extracted from generalized ellipsometry for the $SiO_2$ OAD thin film deposited at α=70°: (a) Porosity gradient. (b) Refractive indexes associated with the three directions (x', y' and z') of the sample reference.

According to this figure, a significant increase in porosity is observed for the bottom part of the deposited layer followed by a slower rise for the remaining thicknesses. This trend is consistent with that observed in SEM (Figure 4.b). The effective refractive indexes resulting from the OAD deposition are therefore highly dependent on the polarization direction and position in the thin film. A difference of 0.11 between $n_{z'}$(z=0) and $n_{x'}$(z=e) can thus be noted. Finally, the thickness for this thin film is obtained equal to e=537 nm and the azimuth angle Φ of the columns in respect with the x axis is close to zero. This last results confirms that the columns axes are aligned to the incident plane (x,z) of the evaporated species.

## 5. Comparison between optical properties and microstructural properties for OAD single layers

The same approach as that described in the previous section (Section 4.) was carried out on a complete series of $SiO_2$ layers deposited by OAD from α=65° to 85° with step of 5°. The morphological information extracted from optical modeling have then been compared with microstructural studies. First, the order of magnitude and evolution of the optical axis angle $\beta_{opt}$ with α are close to the measurements of the column tilts made from cross-sectional SEM observations (Figure 9). We attribute the small discrepancies (maximum difference between the two methods of 8°) to the low refractive index of $SiO_2$ that leads to a small optical anisotropy making the optical simulation not so sensitive to anisotropic parameters.

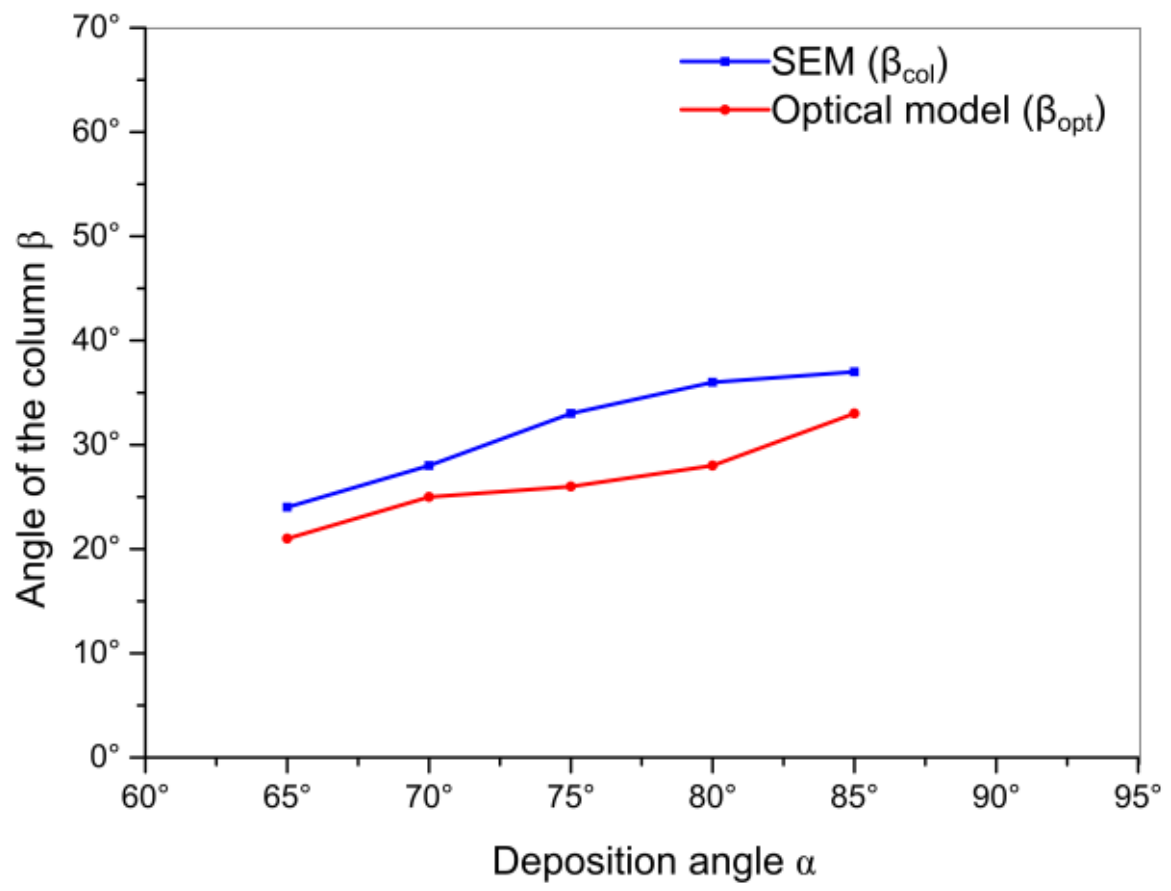

Figure 9 Evolution of the angle β as a function of the deposition angle α determined by optical simulation ($\beta_{opt}$) and by SEM observation ($\beta_{col}$) for a series of $SiO_2$

Another morphological feature that can be extracted from the optical model is the porosity gradient within the different layers in function of the deposition angle α (Figure 10.a). As can be observed, the porosity level in the first growth steps increases faster for high deposition angles. This behavior is consistent with an increase of the competitive character (extinction of the smallest columns) of the growth with α. This evolution of porosity is qualitatively similar to that described by the SEM study added on the figure for the angle of 85°. Here again the low refractive index of $SiO_2$ causes a small influence of the porosity gradient on the optical properties that may explain part of the difference between the model and the microstructural observation. However, we argue that this cannot alone explain this significant discrepancy. We attribute another part of this difference to the nature of the SEM observation that lacks the depth information. The columns all appear to be on the same level on the micrograph, but some could be extinct due to neighbor shadowing but still appear leading to a minimization of the observed porosity level. Furthermore, due to the interaction volume of the electron, the porosity level inside the column isn't perceptible, despite its presence [25], but will affect the optical response. This effect may also contribute to the underestimation of the porosity level from the microstructural study.

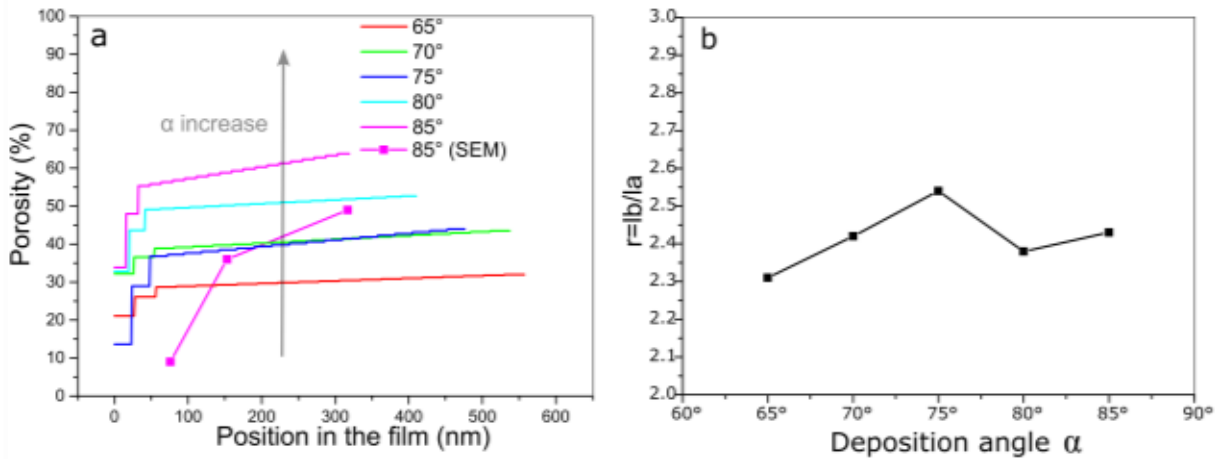

Figure 10 (a) Porosity evolution as a function of z for several deposition angles α, extracted from optical modeling. The estimated surface porosity level extracted from SEM (layer deposited with α=85°) is also shown for comparison. (b) Evolution of the aspect ratio as function of α.

Additionally, from the depolarization coefficients of the optical model, the averaged aspect ratio of the inclusion, r, could be calculated, i.e. considering a medium composed of aligned ellipsoid with the same aspect ratio. The results are presented on Figure 10.b for the angle series. A small evolution of the aspect ratio with α is observed going from 2.30 to a maximal value of 2.54 for the angle of 75°. This tends to indicate a dependency of the column broadening mechanisms with the angle of deposition.

Those values can be compared to the average value of aspect ratio calculated from the microstructural study (Figure 5) and more information on the method of calculation of the average of aspect ratio are given in Supplementary Material 3. Thus, for α=85° an observed microstructural aspect ratio of r=2.15 was found against r=2.43 from the optical model. Here again, a comparable order of magnitude of the aspect ratio is observed between the two methods, although the discrepancy between them could also be due to the weak anisotropy of $SiO_2$. On Supplementary Material 3 the average on the thickness is avoided by considering the depolarizations gradient, and thus an aspect ratio gradient in the model.

## 6. Comparison between optical properties and microstructural properties for an OAD bilayer

From the previous optical models, the effective refractive indices accessible from the different $SiO_2$ OAD layers deposited from α=0° to α=85° have also been determined. By taking into account this information, an antireflective bilayer coating on glass substrate was designed for the 400-1800nm range, and then numerically optimized using the refractive index and thickness of each layers as free parameters to maximize the transmittance. This AR was then experimentally achieved as follows: a $SiO_2$ layer was first deposited by OAD at α=65° with a targeted thickness of 134 nm and refractive index of n=1.30; then, a second $SiO_2$ layer was deposited at α=85° with a targeted thickness of 142 nm and refractive index of n=1.16. An azimuthal rotation of 180° was performed in between the layers in order to minimize any material deposition within the porosity of the first layer during the deposition of the second one.

To evaluate the performances of this bilayer antireflective coating, not only the reflectance (Figure 12.b) but also the transmittance (Figure 12.a) were measured in order to account for absorbance that might arise from the nanostructuration and that could minimize the transparency of the coating. According to Figure 12.a, the prepared bilayer presents high optical performances, in particular a very high and broadband transmittance: average transmittance of 98.97% over [400-1800nm] range. It is also worth noting that the experimental transmittance is close to its original design. This demonstrate that by knowing precisely the effective refractive index of OAD, we were able to obtain the optical properties needed for our designed antireflective coating.

To go further, the previous model (Section 3) was used to describe with more details the optical properties of the bilayer. To make it more reliable, the optical model was constructed in two steps. On the first step, the first layer alone was optically measured and modelled from generalized ellipsometry and spectrophotometry. The obtained optical properties were then conserved, and a second layer was added to simulate the complete bilayer. The fitted parameters of the second layers are the same as described previously for the monolayers with the addition of the angle Φ, azimuthal angle between the first and the second layer, that was allowed to be fitted. The result of this optical simulation are presented on Figure 12.

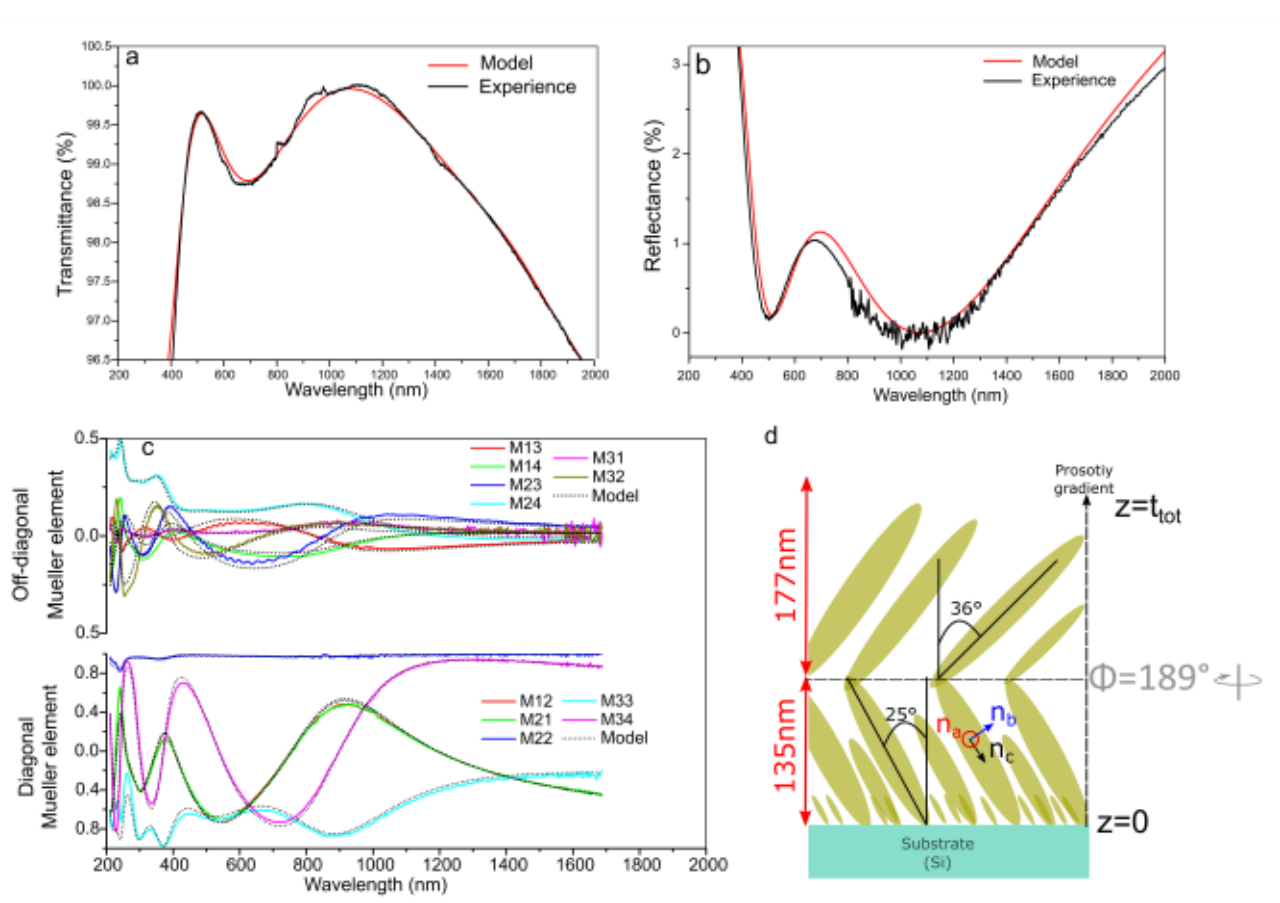

Figure 12 Experimental and simulated (a) transmittance and (b) reflectance spectra of the OAD $SiO_2$ bilayer. (c) Mueller matrix determined by generalized ellipsometry at θ=65° and Φ=45° simulated from the model schematized in (d). The OAD bilayer consists of $SiO_2$ deposited at α=65° and α=85° with an azimuthal rotation between the two layers of Φ=180°

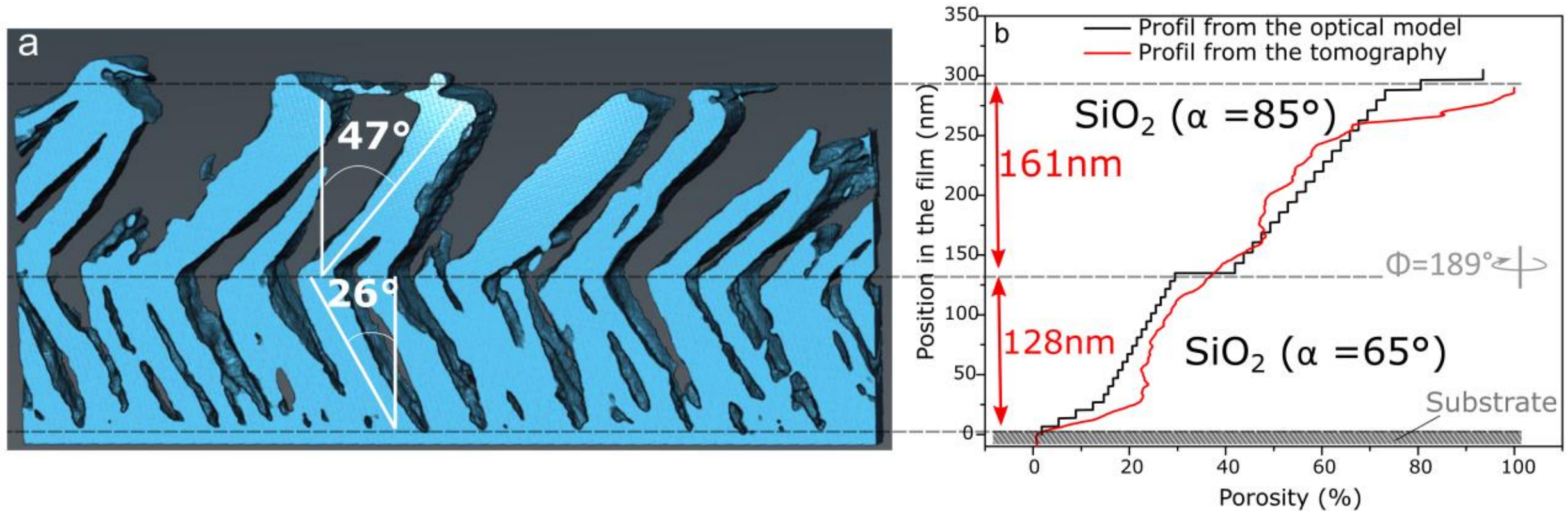

Figure 11 (a) 3D view of the of the reconstructed OAD bilayer. (b) Associated porosity profile as a function of the depth of the film.

It can be noted that the model allows a very precise description of all optical measurements. *Via* the anisotropy information, we determined the column angles $\beta_{opt}$=25° and $\beta_{opt}$=36° respectively for the first and for the second layer. In addition, we were able to demonstrate that the optical axis of the second layer is not aligned in Φ with that of the first layer. An angle of Φ=189° between the first and second layer is noted, which is close to the expected value (Φ=180°). A significant improvement in the mean square error (MSE), about 15%, is achieved by taking into account this parameter. Although anecdotal for stack transmission, the ability to access this value demonstrates the high sensitivity of generalized ellipsometry. In practice, this slight deviation can be explained by a slight disorientation of the sample during its placement at Φ=180° for the deposition of the second layer compared to the deposition of the first, this alignment being done approximately by the visual inspection.

To get a precise comparison between optical properties and microstructure, a 3D morphological analysis of the bilayer has been performed at the nanoscale from HAADF-STEM electron tomography experiments reported in our previous paper [26]. After data reconstruction, this characterization enabled us to have a precise volumetric information on the layer, especially on the porosity gradient along the depth of the film as presented on the Figure 11.

Interestingly, the measured thicknesses of the first and second layers obtained from the 3D reconstruction (128 nm and 161 nm, respectively) are quite close to the ones extracted from the optical model (135 nm and 177 nm, respectively). Furthermore, the column angle in the first layer determined from electron tomography ($\beta_{col}$=26°) matches the one expected from the optical model ($\beta_{opt}$=25°). It should be noted however that the column angle of the second layer determined from the model, $\beta_{opt}$=26°, is quite different from the one observe, $\beta_{col}$=47°. This difference for the second layer might come as it was previously stated from the weak optical anisotropy of $SiO_2$. Interestingly this value is significantly higher than the value found for the same angle of deposition but grown on plane silicon substrate ($\beta_{col}$=37°). This highlights a difference of the growth behavior of the OAD layer when deposited on prepatterned thin film as it was previously demonstrated [27].

Beside the extraction of some basic morphological information of the OAD coatings, like thicknesses and column tilts, electron tomography appears here as a powerful approach for further quantitative extraction of the porosity gradient within the full layer (Figure 11.b). The very nice agreement of this porosity profile compared to that extracted from advanced optical simulations demonstrates the validity of the optical model used in this work, and indicate that it is possible to reliably describe the porosity profile from optical characterization. This type of information is crucial for further optical design optimization, by considering the porosity gradient naturally present in OAD layer to optimize optical function from such layer as antireflection for example.

## IV. CONCLUSIONS

In this work, we demonstrated by microstructural study that the growth process of OAD will produce nanocolumns with an anisotropic broadening of the section and an increasing porosity rate for an increasing thickness. These morphological aspects have an impact on optical properties that needs to be considered for practical applications. This was measured using generalized ellipsometry and spectrophotometry. Guided by the microstructural study, we established an optical model using the ABEMA combined with a gradient of porosity. This model describes with great accuracy both the generalized ellipsometry and spectrophotometry measurements. The parameters extracted from this model like the column angle, porosity rate and aspect ratio of the columns are close to the ones observed from the microstructural study which validate the merits of our model. Knowing with high accuracy the optical properties of OAD thin films allowed us to optimize precisely a $SiO_2$ antireflective bilayer that shows a very high level of transmittance. Finally, using electron tomography, we also confirmed that the optical model describes closely the morphological properties of the bilayer making it an interesting tool to get easy access to morphological parameters from optical measurements.

## ACKNOWLEDGEMENTS

This work was supported by the DGA (Direction Générale de l’Armement), the French Defense Procurement Agency. This work has been partially supported by « Nouvelle Aquitaine » Region and by European Structural and Investment Funds (ERDF reference P-2016-BAFE-209): IMATOP project.
A. J. Santos thanks the financial support of the IMEYMAT Institute and the Spanish Ministerio de Educación y Cultura for the concessions of grants (ICARO-173873 and FPU16-04386).
The “Talent Attraction Program” of the University of Cádiz is also acknowledged by supporting B. Lacroix contract code E-11-2017-0117214.
This work was carried out in the framework of the associate laboratory PRIMEO ("Partnership for Research and Innovation in Emerging Materials for phOtonics") between Safran Electronics & Defense and Pprime Institute.

# Supplementary material 1

This supplementary material is a python code presented in the form of a Jupyter notebook named ‘Optical simulation of SiO2 anisotropic layer using generalized ellipsometry.ipynb’ accessible at https://github.com/Quikim/Optic_lab. The interactive content of this notebook can directly be launched using Binder under the following link: https://mybinder.org/v2/gh/Quikim/Optic_lab/master. An archive of this code is also available at Zenodo with the following DOI: 10.5281/zenodo.3576393

# Supplementary material 2

On Figure 14 are presented the SEM micrographs before and after the image treatment described in the experimental section. On the processed micrographs the best fit ellipses estimating the columns agglomerates are presented in red and each column agglomerate has a unique color.

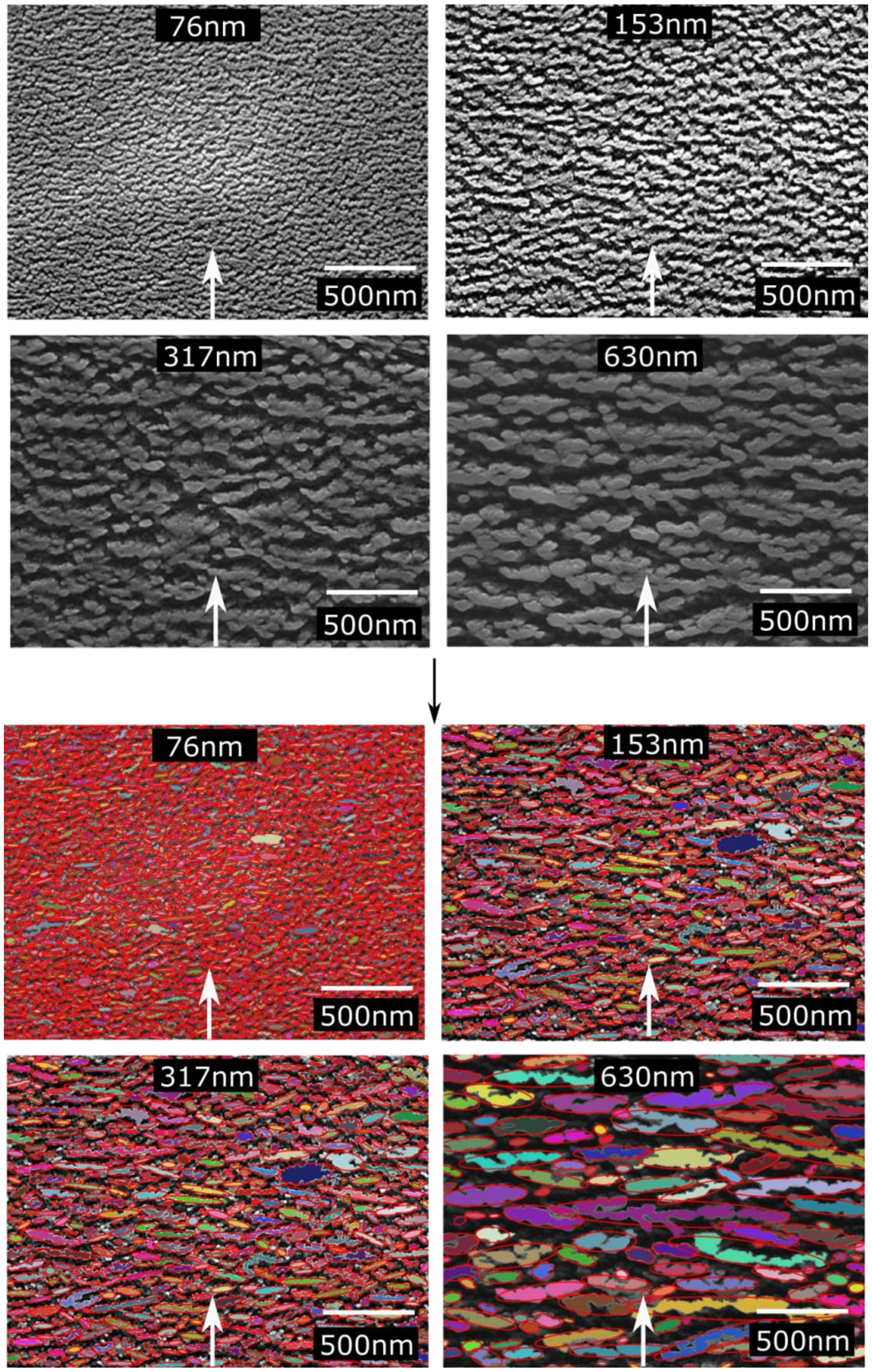

Figure 13 SEM micrographs before and after image processing in plane view of a $SiO_2$ sample deposited by OAD at α=85° for different thicknesses. The arrow indicate the direction of the deposited species

# Supplementary material 3

## Calculation of the mean aspect ratio

The optical model developed in our article allows to quantify the aspect ratio representative to the morphology of the sample. To compare these values with the microstructural study, for the sake of simplicity, we considered the center of the nanocolumns to be distributed homogenously on the sample. The mean aspect ratio characterizing the morphological anisotropy of the sample is then solely due to the orientation and ellipticity of the structures. To calculate the mean aspect ratio, the orientation of the ellipses approximating the nanocolumns, has to be projected on the horizontal axis as follow:

$$\varepsilon_j = \frac{\sum_i la_i \cos\gamma_i + lb_i \sin\gamma_i}{\sum_i lb_i \cos\gamma_i + la_i \sin\gamma_i}$$

Where $\varepsilon_j$ is the eccentricity at the thickness j, $\gamma_i$ is the angle of the ellipse i with the horizontal axis and $la_i$ and $lb_i$ are respectively the long and short axis of the ellipse i. Since in the optical model, as a first approach, the anisotropy is considered homogenous for the thickness, the mean value from the thickness of 76nm to 355nm is then calculated.

## Taking into account an anisotropy gradient

As a first approach the model presented in the paper considered constant depolarization coefficient with thickness. Here we present the result of the model implementing graded depolarization coefficients. The depolarization coefficients La, Lb and Lc provide information on the evolution of the shape factor characteristic of the OAD layer with thickness (Figure 14.a). The Lc coefficient is very low, which indicates very elongated columns with a low cross-section. Lc varying very slightly in thickness (Lc(z)≈0), only the variation of La and Lb is considered. The aspect ratio of the ellipses ε determined to simplify from the coefficients La and Lb increases with thickness, which reflects an increasingly anisotropic morphology. This ratio increases sharply for the early stages of growth. On the other hand, Figure 14.b shows an increasingly rapid increase in the aspect ratio with the angle of α. This confirms that for the same value of z the aspect ratio increases with the deposition angle.

We recall here that the SEM study determined an aspect ratio evolving from ε=3 to ε=3.5 for a thickness varying from 76nm to 317nm on the deposit made at α=85° that was added on Figure 14.b. The value of the aspect ratio and the optically determined evolution follow the same trend. This evolution of the aspect ratio of the ellipse with z is in agreement with that described by C Lopez-Santos et al.[4]

Finally, it should be noted that taking into account the gradient of the depolarization coefficients in thickness only partially improves the quality of the optical simulation (5% gain in the MSE value). These last results should be confirmed by other characterization means to conclude that the microscopic characterizations derived from the optical model are sensitive to these tenuous evolutions.

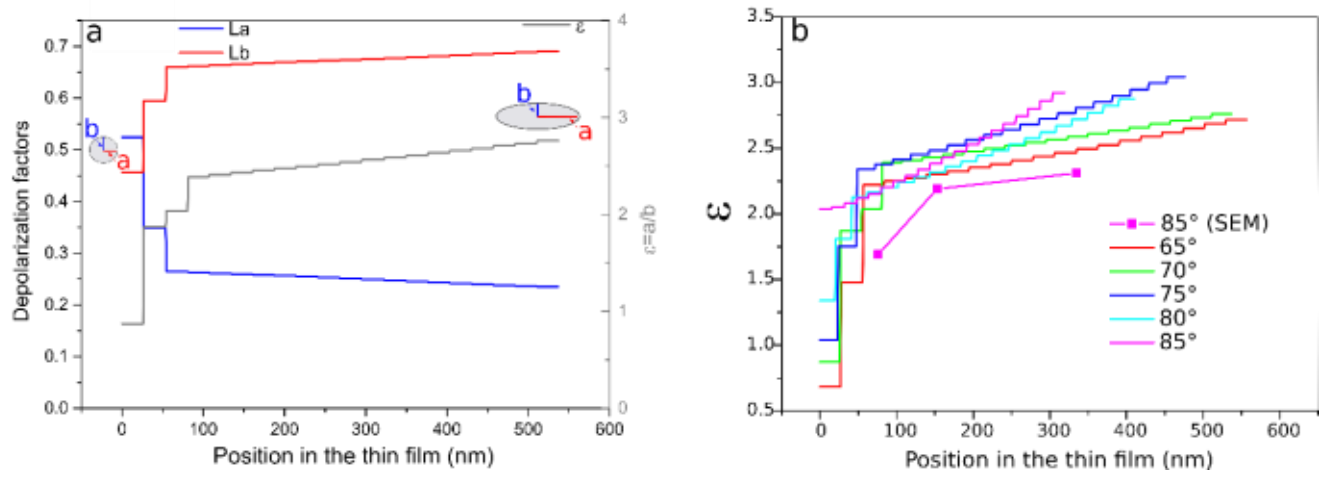

Figure 14 (a) Evolution of depolarization coefficients as a function of the position in the film for a $SiO_2$ sample deposited at α=70° (b) evolution of the aspect ratio ε as a function of the position in the film for several deposition angles of the same SiO2 series. The aspect ratio values determined by SEM are attached in (b)

# Supplementary material 4

To demonstrate the need to include optical losses when simulating an OAD layer, on Figure 14 is presented the optical losses (defined by 1-R-T) for an OAD layer of SiO2 deposited at α=85° with a thickness of 630nm. The simulated absorption is added assuming the material constituting the $SiO_2$ has the same extinction coefficient as the dense $SiO_2$ reference (Simulated absorption). The optical absorption coming from this extinction coefficient is too low to explain the optical losses measured on the OAD sample. We attribute this increase of light attenuation to scattering for two reasons. First, considering that it comes from absorption would mean that the OAD nanostructuration introduces a pollutant that absorbs light in the visible range as $SiO_2$ is highly transparent in this range. However, the presence of such pollutant wasn't detected through EDX. But also more importantly, we have demonstrated through FDTD calculation, presented in a previous article, that due to the size of the nanocolumns as they broaden, scattering will occur as their dimension isn't small enough in regard of the incoming wavelength [26]. The optical losses observed experimentally for short wavelengths therefore do not result from absorption, unlike those observed in the near infrared resulting from low substrate absorption. Thus, for the optical model to be complete the simulation of optical losses needs to be done. In this article we have done it including the Urbach light attenuation model as this allows the measurements to be reproduced accurately Figure 14.

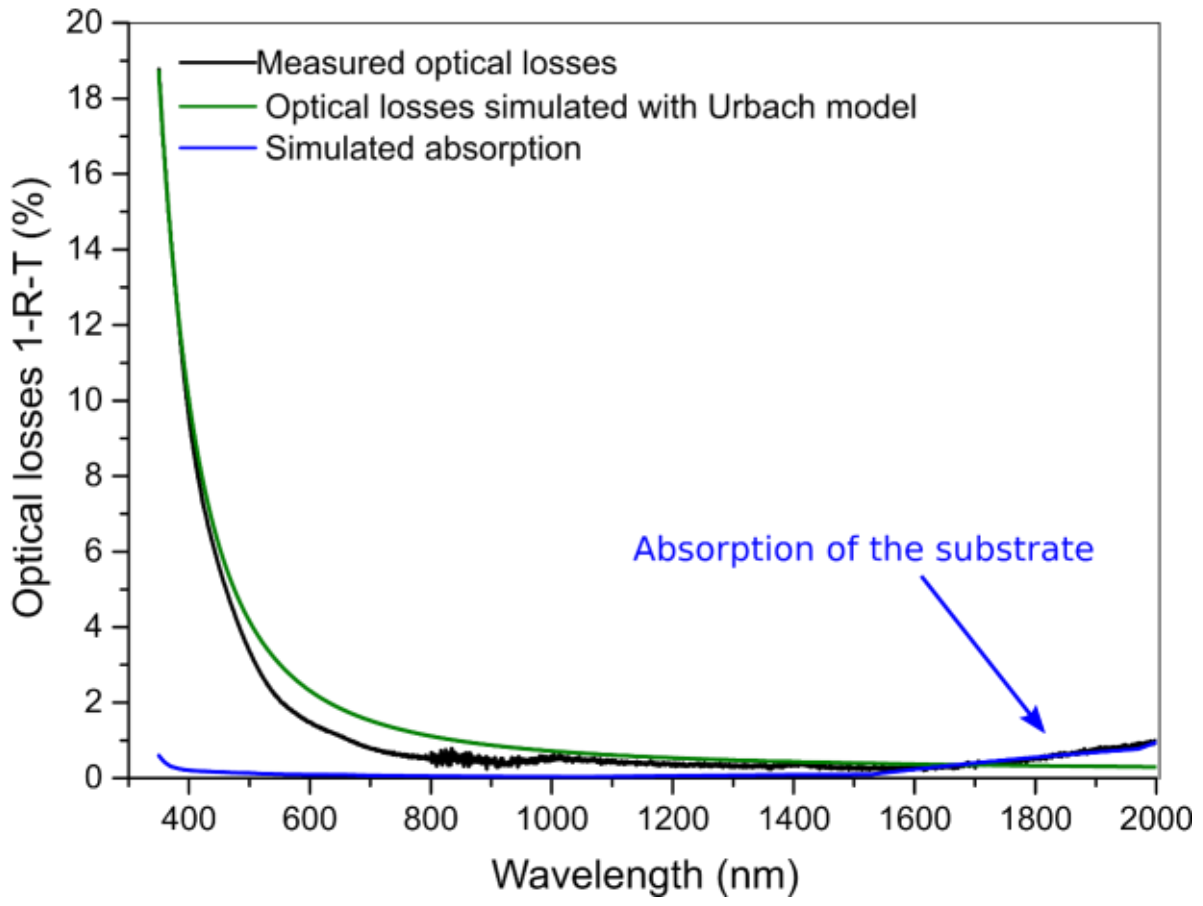

Figure 15 Optical losses measured or calculated on a sample of SiO2 deposited at α=85°: optical losses (scattering + absorption) extracted from spectrophotometry measurements, experimental losses extracted from the Urbach model (diffusion + absorption). For reference, losses related to a dense layer of $SiO_2$ on a glass substrate